\documentclass[final]{IEEEtran}
\usepackage{amsthm,amssymb,graphicx,multirow,amsmath,color,amsfonts}
\usepackage[update,prepend]{epstopdf}
\usepackage[noadjust]{cite}
\usepackage[latin1,utf8]{inputenc}
\usepackage{tikz}
\usepackage{bbm} 
\usepackage{pdfpages}
\usepackage{tabulary}
\usepackage{multirow}
\usepackage{comment}
\usepackage{subfigure}
\usepackage{float}
\usepackage{mathtools,amssymb,lipsum}
\usepackage{cuted}
\def\nb0{{\mathbf{0}}}
\def\nb1{{\mathbf{1}}}

\newtheorem{lemma}{Lemma}

\newtheorem{definition}{Definition}

\newtheorem{remark}{Remark}

\allowdisplaybreaks 
\begin{document}
\title{On the Uplink SINR Meta Distribution of UAV-assisted Wireless Networks}
\author{
	Yujie Qin, Mustafa A. Kishk, {\em Member, IEEE}, and Mohamed-Slim Alouini, {\em Fellow, IEEE}
	\thanks{Yujie Qin and Mohamed-Slim Alouini are with Computer, Electrical and Mathematical Sciences and Engineering (CEMSE) Division, King Abdullah University of Science and Technology (KAUST), Thuwal, 23955-6900, Saudi Arabia
		Arabia. Mustafa Kishk is with the Department of Electronic Engineering, Maynooth University, Maynooth, W23 F2H6, Ireland. (e-mail: yujie.qin@kaust.edu.sa; mustafa.kishk@mu.ie; slim.alouini@kaust.edu.sa).} 
	
}
\date{\today}
\maketitle
\begin{abstract}
This letter studies the  signal-to-interference-plus-noise (SINR) meta distribution of uplink transmission of UAV-enabled wireless networks with inversion power control. Within a framework of stochastic geometry, the Matern cluster process (MCP) is used to model the locations of users and UAVs. Conditional success probability and moments are derived to compute the exact expression and moment matching approximation of SINR meta distribution (beta approximation). Specifically, the effect of the power control compensation factor and UAV altitude are studied. Our numerical results show that UAV altitude has a higher impact on the system reliability than transmit power at low values of SINR threshold since users benefit more from establishing line-of-sight (LoS) links with UAVs.
\end{abstract}
\section{Introduction}
The performance of wireless systems depends heavily on the spatial locations of users and base stations (BSs). Stochastic geometry provides mathematical models and methods to analyze and obtain design insights of the wireless networks with randomly placed nodes \cite{haenggi2015meta,elsawy2017meta}. While most of the stochastic geometry-based literature on UAV-enabled networks
 are confined to the spatial average, the performance of individual links draws less attention \cite{elsawy2016modeling}. On the side of operators, it is crucial to obtain key information about "the distribution of success probability of the individual link in a given network" \cite{kalamkar2019simple}, which reveals the reliability and quality of service (QoS) of the network. This system insight is called signal-to-interference-ratio/signal-to-interference-plus-noise (SIR/SINR) meta distribution \cite{haenggi2015meta},  defined as a complementary cumulative distribution function (CCDF) of the conditional success probability and provide  the information on each link instead of a spatial average of the whole network \cite{haenggi2021meta1,haenggi2021meta2}.  

The detailed definition of meta distribution and examples, PPP and Poisson bipolar networks, are provided in \cite{haenggi2021meta1}.
Lots of works on the meta distribution has been done in Poisson and non-Poisson cellular networks. For instance, authors in  \cite{kouzayha2021meta} analyzed the  Poisson binomial point process (BPP)-based networks, and \cite{saha2020meta,kalamkar2019simple} utilized the deployment gain to approximate the SIR meta distribution for general networks. Since computing the SIR meta distribution requires calculating the moments of conditional success probability, the exact equation is hard to derive. The beta approximation is a commonly used approximation for SIR meta distribution, which only requires the first two moments of the conditional success probability \cite{haenggi2015meta}. Fourier-Jacobi expansion of moments to reconstruct meta distribution is used in \cite{8421026}. The author in \cite{haenggi2021meta2} provided a close-form of SIR meta distribution which only requires the nearest interfer and is an upper bound of system performance. The separable region of SIR meta distribution for any independent fading is analyzed in \cite{9072358}.

 Different from the existing literature, which mainly focus on the spatial average performance of UAV-enabled networks, this letter studies the SINR meta distribution of uplink transmission of UAV-enabled networks with inversion power control. Unlike our previous work \cite{Mbref}, this letter studies the uplink transmission, in which the transmit power is a function of transmission distance, hence, the results of \cite{Mbref} cannot be used in the uplink scenario directly.  We study the influence of UAV altitude and power control compensation factors on the system performance. The results reveal several system insights, such as the reliability of the system significant improvement by increasing the UAV altitudes, hence, increasing the probability of establishing line-of-sight (LoS) link with serving BSs.

\section{System Model}
In this letter, we analyze the SINR meta distribution of the uplink transmission of a UAV-enabled wireless network which is composed of UAVs and users. In this work, the users are cluster distributed and modeled by the Matern cluster process (MCP) \cite{9153823}. It means that locations of user cluster centers are spatially distributed according to a homogeneous Poisson point process (PPP) $\Phi_b$ with density $\lambda_b$, and users are uniformly distributed in the cluster. UAVs are assumed to hover above the user cluster centers at a fixed altitude $h$, and  the clusters are with radii $r_c$. For simplicity, we assume that users associate with their cluster UAVs, and in each resource block, only one user can communicate with the cluster UAV. Let $\Phi_u$ with density $\lambda_u$ be the point set of the locations of the active users. Since we assume only one user is active at each resource block, the density of active users  $\lambda_u = \lambda_b$.

When the user communicates with its cluster UAV,  a fractional path-loss inversion power control with compensation factor $\epsilon_{\{l,n\}}$ is used, where the subscripts $l,n$ denotes the LoS and NLoS link with the cluster UAV, respectively.
Besides, a standard path-loss model with exponents $\alpha_{\{l,n\}}$ and  Nakagami-m channel fading model are used in this work.

\subsection{Inversion Power Control}
 In this letter, we condition a UAV to be located at the origin and focus on the SINR of this UAV. It becomes the reference UAV on averaging over the point process and is equivalent to any other arbitrary deterministic location owing to the stationarity of PPP. Consequently, the user served by reference UAV and their established link become the reference user, $u_o$, and reference link, respectively.
 
Assume that inversion power control technology is used in the uplink transmission \cite{elsawy2017meta,6786498}. This technique is widely used to increase the transmit power of users to compensate for the path loss \cite[Chapter 4]{goldsmith2005wireless} and we consider the transmit power to be truncated due to the limited power of a user equipment.  Therefore, the transmit power of a user depends on established NLoS/LoS channel and transmission distance, is given by
 \begin{align}
 	p_{t} = \left\{
 	\begin{aligned}
 		p_{t, l} &=  \rho_l R_{u}^{\alpha_{l}\epsilon_{l}}, \text{in the case of LoS},\\
 		p_{t, n} &=  \rho_n R_{u}^{\alpha_{n}\epsilon_{n}}, \text{in the case of NLoS},\\
 	\end{aligned}
 	\right.
 \end{align}
where $R_u$ is the Euclidean distance between the user and the serving UAV, $\rho_{\{l,n\}}$ is a power control parameter to adjust the received power at the serving BS, $\epsilon_{l,n}$ is the compensation factor,  $\eta_{ l}$ and $\eta_{ n}$ denote the mean additional losses and $\alpha_{\{l,n\}}$ are the path-loss exponents for LoS and NLoS transmissions, respectively. Assuming that the maximum transmit power is $p_u$, the range of $\epsilon_{l,n}$ is
\begin{align}
	0\leq\epsilon_{l,n} \leq  \frac{1}{\alpha_{\{l,n\}}}\log_{\sqrt{r_c^2+h^2}}\bigg(\frac{p_u}{\rho_{\{l,n\}}}\bigg),
\end{align} 
in which $r_c$ is the user cluster radii, $h$ is the altitude of the UAV. Consequently, the received power at the UAV is
\begin{align}
	p_{r} = \left\{
	\begin{aligned}
		p_{r, l} &= \eta_{l}\rho_l G_{l}R_{u}^{-(1-\epsilon_l)\alpha_{l}}, \text{LoS link},\\
		p_{r, n} &= \eta_{n}\rho_n G_{n}R_{u}^{-(1-\epsilon_n)\alpha_{n}}, \text{NLoS link},\\
	\end{aligned}
	\right.
\end{align}
in which  $G_{ l}$ and $G_{ n}$ present the fading gains and follow Gamma distribution with shape and scale parameters ($m_{ l},\frac{1}{m_{ l}}$) and ($m_{ n},\frac{1}{m_{ n}}$) for LoS and NLoS scenarios, respectively. Given the horizontal distance $r$, the occurrence probability of a LoS link established between the serving UAV and the user  is given in ~\cite{al2014optimal} as
\begin{align}
	P_{ l}(r) & =  \frac{1}{1+a \exp(-b(\frac{180}{\pi}\arctan(\frac{h}{r})-a))},\label{pl_pn}
\end{align}
where $a$ and $b$ are two environment variables. Consequently, the probability of NLoS link is $P_{ n}(r)=1-P_{  l}(r)$.

\subsection{The SINR Meta Distribution}
We are interested in analyzing the uplink reliability of a UAV-assisted network. To do so, we study the SINR meta distribution of UAV-assisted networks, which is defined as the fraction of users exceeding a predefined SINR threshold $\theta$ of a network. For an arbitrary realization of $\Phi_b$ and $\Phi_u$, the uplink SINR meta distribution is the percentiles of users that achieve uplink coverage.
\begin{definition}[SINR Meta Distribution] The SINR meta distribution of uplink is defined as
	\begin{align}
	\bar{F}_{P_s}(\theta,\gamma) = \mathbb{P}(P_s(\theta) > \gamma),\label{eq_MetaF}
\end{align}
where $\gamma \in [0,1]$ and $P_s(\theta)$ is the conditional success probability (conditioned on the realizations of $\Phi_b$ and $\Phi_{u}$) given by,
\begin{align}
	P_s(\theta) &= \mathbb{P}({\rm SINR}>\theta)= \mathbb{P}\bigg(\frac{p_r}{I+\sigma^2}>\theta\bigg),
\end{align}
in which $\sigma^2$ is the noise power and $I$ is the interference.
\end{definition}

\section{Meta Distribution Analysis}
In this section, we provide the analysis for the conditional success probability and SINR meta distribution of the  uplink transmission of  a UAV-assisted network.  Recall that the reference UAV is assumed to serve the users in its cluster, the distance between the cluster UAV and the user is $R_u$. The probability density function of $R_u$, denoted by $f_{ R_u}(r)$, is
	\begin{align}
		f_{ R_{u}}(r) &= \frac{2r}{r_{c}^2}, \quad h\leq r \leq \sqrt{r_c^2+h^2}.
	\end{align}

\subsection{Success Probability}
Recall that for a fixed realization, we randomly select one user as our reference user, the conditional success probability of the reference link, which is conditioned on the realization of $\Phi_{u}$,  is given by
\begin{align}
	P_s(\theta) \stackrel{(a)}{=}& P_{s, l}(\theta)+P_{s, n}(\theta)\nonumber\\
=& P_l(R_u)\mathbb{P}\bigg(\frac{p_{t,l}\eta_{ l} G_l R_u^{-\alpha_l}}{I+\sigma^2}>\theta|{\rm LoS}\bigg)\nonumber\\
&+P_n(R_u)\mathbb{P}\bigg(\frac{p_{t,n}\eta_{ n} G_n R_u^{-\alpha_n}}{I+\sigma^2}>\theta|{\rm NLoS}\bigg),
	\end{align}
in which $P_{s,l}(\theta)$ and $P_{s,n}(\theta)$ denote the probability conditioned on the establish link is LoS or NLoS, and step (a) follows from the law of total probability,
\begin{align}
&I = \sum_{u_i\in\Phi_{u}^{ll}}\rho_l\eta_{ l} G_l R_{u_i}^{\alpha_{l}\epsilon_l}D_{u_i}^{-\alpha_{l}}+\sum_{u_i\in\Phi_{u}^{nl}}\rho_n\eta_{l} G_l R_{u_i}^{\alpha_{n}\epsilon_n}D_{u_i}^{-\alpha_{l}}\nonumber\\
&+\sum_{u_i\in\Phi_{u}^{ln}}\rho_l\eta_{ n} G_n R_{u_i}^{\alpha_{l}\epsilon_l}D_{u_i}^{-\alpha_{n}}+\sum_{u_i\in\Phi_{u}^{nn}}\rho_n\eta_{n} G_n R_{u_i}^{\alpha_{n}\epsilon_n}D_{u_i}^{-\alpha_{n}},\label{eq_I}
\end{align}
in which $\Phi_{u}^{ll}$, $\Phi_{u}^{ln}$, $\Phi_{u}^{nl}$ and $\Phi_{u}^{nn}$ are subsets of $\Phi_{u}\setminus u_o$ (recall that $u_o$ denotes the location of the reference user) denote the locations of interfering users which establish LoS/NLoS links with their cluster UAVs and have LoS/NLoS links with the reference UAV, respectively, $D_{\{\cdot\}}$ denotes the distances between the interfering users and the reference UAV, and $R_{ui}$ denotes the distances between the interfering UAVs and their serving users.
\begin{lemma}[Conditional Success Probability]\label{lemma_ps}
	 The conditional success probability in the case of LoS and NLoS transmission with the serving UAV is
	\begin{align}
	&	P_{s,l}(\theta) \approx P_l(R_u)\sum_{k = 1}^{m_l}\binom{m_l}{k}(-1)^{k+1}\exp(-g_l(R_u)\sigma^2)\nonumber\\
		&\prod_{u_i\in\Phi_{u}^{e,j}}f(e,j,g_{l}(R_u),R_{u_i},D_{u_i}),\quad \{e,j\}\in\{l,n\},\label{eq_ps_l}\\
	&	P_{s,n}(\theta) \approx P_n(R_u)\sum_{k = 1}^{m_n}\binom{m_n}{k}(-1)^{k+1}\exp(-g_n(R_u)\sigma^2)\nonumber\\
		&\prod_{u_i\in\Phi_{u}^{e,j}}f(e,j,g_{n}(R_u),R_{u_i},D_{u_i}),\label{eq_ps_n} \quad \{e,j\}\in\{l,n\},
	\end{align}
	in which the approximation signs come from the use of the upper bound of Gamma distribution and
\begin{small}
		\begin{align}
		f(e,j,g_{\{l,n\}}(r),R_{u_i},D_{u_i}) = \bigg(\frac{m_j}{m_j+g_{\{l,n\}}(r)\rho_e\eta_{ j} R_{u_i}^{\alpha_{e}\epsilon_e}D_{u_i}^{-\alpha_{j}}}\bigg)^{m_j},
	\end{align}
\end{small}
where $g_l(r) = k\beta_2(m_l)m_l\theta r^{(1-\epsilon_{l})\alpha_{ l}}(\rho_l\eta_{ l})^{-1}$, $g_n(r) = k\beta_2(m_n)m_n\theta r^{(1-\epsilon_{n})\alpha_{n}}(\rho_n\eta_{n})^{-1}$, and $\beta_2(m) = (m!)^{-1/m}$.
\end{lemma}
\begin{IEEEproof}
	See Appendix \ref{app_proof_ps}.
\end{IEEEproof}

\subsection{$b$-th Moments}
The $b$-th moment of the conditional success probability is derived	by taking the expectation over the locations of the interfering users and distance to the serving BS,	given by
\begin{align}
	M_b(\theta) &= 	M_{b,l}(\theta)+ M_{b,n}(\theta) \nonumber\\
	&=\mathbb{E}[P_{s, l}^{b}(\theta)]+\mathbb{E}[P_{s, n}^{b}(\theta)].
\end{align}
Therefore, the $b$-th moments of the network is given in the following lemma.

\begin{lemma}[$b$-th Moments] \label{lemm_Mb}The $b$-th moments of the uplink transmission of a UAV-assisted network  is
\begin{align}
	M_{b,l}(\theta) \approx&  \sum_{k_1 = 1}^{m_l}\cdots\sum_{k_b = 1}^{m_l}\binom{m_l}{k}\cdots\binom{m_l}{k_b}(-1)^{k_1\cdots k_b+b}\nonumber\\
	& \int_{0}^{r_c}P_l(r)\mathcal{L}(g_{l,1}(r),\cdots,g_{l,b}(r))f_{ R_u}(r){\rm d}r,\nonumber\\
	M_{b,n}(\theta) \approx& \sum_{k_1 = 1}^{m_n}\cdots\sum_{k_b = 1}^{m_n}\binom{m_n}{k}\cdots\binom{m_n}{k_b}(-1)^{k_1\cdots k_b+b}\nonumber\\
	& \int_{0}^{r_c}P_n(r)\mathcal{L}(g_{n,1}(r),\cdots,g_{n,b}(r))f_{ R_u}(r){\rm d}r,\label{eq_Mb}
\end{align}
where $g_{l,i}(r) = k_i\beta_2(m_l)m_l\theta r^{(1-\epsilon_{l})\alpha_{ l}}(\rho_l\eta_{ l})^{-1}$,  $g_{n,i}(r) = k_i\beta_2(m_n)m_n\theta r^{(1-\epsilon_{n})\alpha_{ n}}(\rho_n\eta_{ n})^{-1}$, and $\mathcal{L}(g_{1}(r),\cdots,g_{b}(r))$ is shown in Appendix \ref{app_laplace}.
\end{lemma}
\begin{IEEEproof}
	(\ref{eq_Mb}) is derived using the probability generation functional (PGFL) of PPP, moment generating function (MGF) of Gamma distribution, PDF of $R_u$, and \cite[Theorem 1]{Mbref}.
\end{IEEEproof}

As mentioned in  \cite{haenggi2015meta} and \cite{gil1951note}, the meta distribution (\ref{eq_MetaF}) is solved by using the Gil-Pelaez theorem and the exact expression is given by
\begin{align}
	\bar{F}_{P_s}(\theta,\gamma) &=  \frac{1}{2}+\frac{1}{\pi}\int_{0}^{\infty}\frac{\Im(\exp(-jt\log \gamma)M_{jt}(\theta))}{t}{\rm d}t,\label{eq_MetaExact}
\end{align}
where $j$ is the imaginary unit, $M_{jt}(\theta)$ is the ${jt}$-th moment of $P_s(\theta)$, computed by replacing the $b$ in the $b$-th moment of conditional success probability $M_b(\theta)$, and $\Im(\cdot)$ is the imaginary part of a complex number.

Generally, computing the meta distribution using (\ref{eq_MetaExact}) is difficult since it requires computing the imaginary moments, which is difficult and not practical. Alternatively, we use the beta approximation, which only requires to compute the first and the second moments and shows great matching \cite{haenggi2015meta}.
\begin{remark}[Beta Approximation]
In the case of the beta approximation, the SINR meta distribution can be approximated, as mentioned in \cite{haenggi2021meta1,haenggi2021meta2}, by
\begin{align}
	\bar{F}^{'}_{P_s}(\theta,\gamma)	\approx 1-&I_{\gamma}(\frac{M_1(\theta)(M_1(\theta)-M_2(\theta))}{M_2(\theta)-M_1^2(\theta)},\nonumber\\
	& \frac{(M_1(\theta)-M_2(\theta))(1-M_1(\theta))}{M_2(\theta)-M_1^2(\theta)}),\label{eq_MetaBetaApp}
\end{align}
where $
	I_x(a,b)= \frac{\int_{0}^{x}t^{a-1}(1-t)^{b-1}{\rm d}t}{B(a,b)}$, and $
	B(a,b) = \int_{0}^{1}t^{a-1}(1-t)^{b-1}{\rm d}t$.
\end{remark}

\section{Numerical Results}
In this section, we validate the theoretical results via Monte Carlo simulations. To ensure the accuracy, we run a large number of iterations $10^{6}$. For the given system model, we first generate a PPP for the locations of user cluster centers and UAVs are located at the altitude $h$ above the user cluster centers. We then generate the locations of users, which are uniformly distributed within the user clusters. While the realizations of all the locations are fixed, the fading realizations change in each iteration, and we compute the success probability for each user link. Finally, we obtain the CCDF of SINR of the given realization.  Unless stated otherwise, we use the system parameters listed herein Table \ref{par_val}.
\begin{table}[h]\caption{Table of Parameters}\label{par_val}
		\vspace{-4mm}
	\centering
	\begin{center}
		\resizebox{1\columnwidth}{!}{
			\renewcommand{\arraystretch}{1}
			\begin{tabular}{ {c} | {c} | {c}  }
				\hline
				\hline
				\textbf{Parameter} & \textbf{Symbol} & \textbf{Simulation Value}  \\ \hline
				Density of UAVs, active users & $\lambda_{ u}$, $\lambda_{ b}$ & $1$ km$^{-2}$, $1$ km$^{-2}$ \\\hline
				MCP disk Radius & $r_c$ & 100 m \\\hline
				Environment parameters (dense area) & $ (a,b)$ & $(12,0.11)$ \\\hline
				Power control parameters & $\rho_{\{l,n\}}$ & 0.01 W, 0.01 W\\\hline
	            Maximum transmission power &  $p_{u}$ &  2 W\\\hline
				Noise power & $\sigma^2 $ & $10^{-9}$ W\\\hline
				N/LoS path-loss exponent & $\alpha_{ n},\alpha_{ l}$ & $4,2.1$ \\\hline
				N/LoS fading gain & $m_{ n},m_{ l}$ & $1,3$ \\\hline
				N/LoS additional loss& $\eta_{ n},\eta_{ l}$ & $-20,0$ dB 
				\\\hline\hline
		\end{tabular}}
	\end{center}
	\vspace{-4mm}
\end{table}

\begin{figure}[h]
	\centering
	\subfigure[]{\includegraphics[width=0.8\columnwidth]{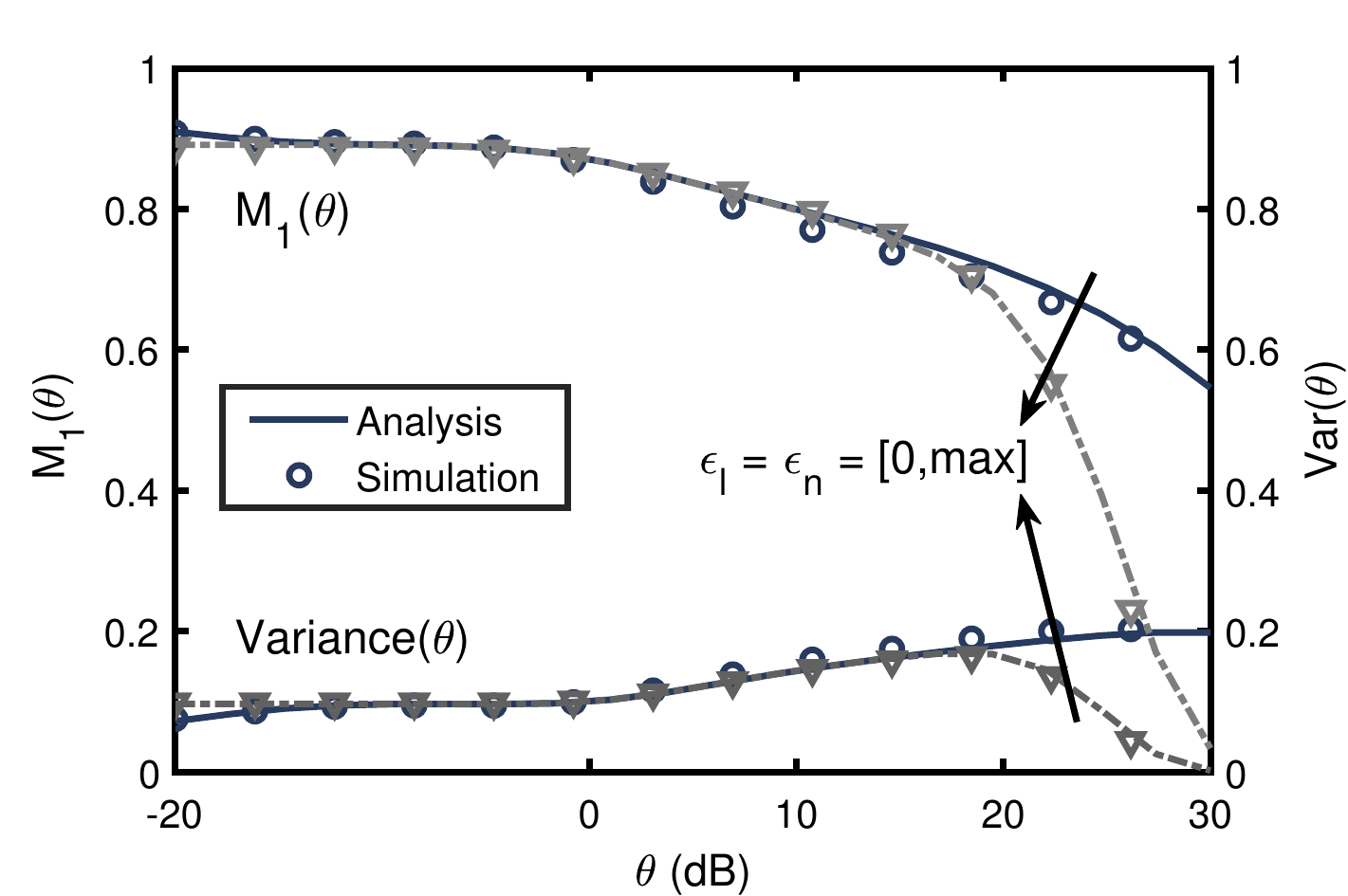}}
	\subfigure[]{\includegraphics[width=0.8\columnwidth]{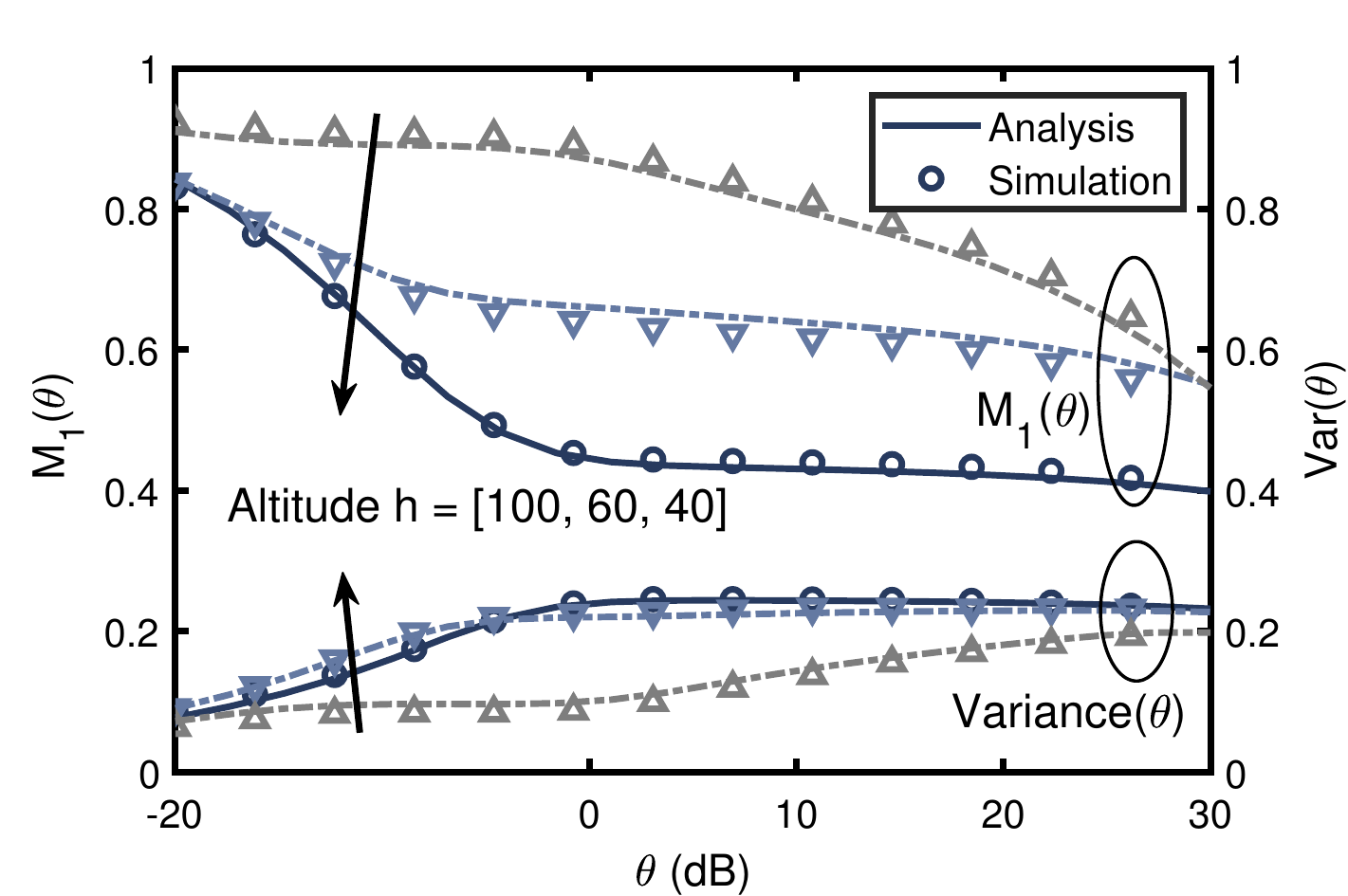}}
	\caption{The simulation and analysis results of the first moment  and the variance of the  system under \textbf{(a)} different compensation factor and $h = 100$ m, and \textbf{(b)} different $h$ and $\epsilon_{l,n} = \max$.}
	\label{fig_M1M2}
\end{figure}
	\vspace{-1mm}

We first validate the analytical expressions by plotting the first and the second moments of the conditional success probability. Fig. \ref{fig_M1M2} shows the uplink standard success probability $M_1(\theta) = \mathbb{E}[P_s(\theta)]$ and the variance of conditional success probability ${\rm Var}(\theta) = M_2(\theta)-M_1^2(\theta)$ as a function of $\theta$ for  uplink  transmission of the UAV-assisted cellular network under different compensation factors and three altitudes.  Note that the first moment of conditional success probability is also known as the coverage probability, which is a spatial average performance of the proposed network.  Interestingly, the value of $\epsilon_{l,n}$ only influences the system performance at large values of $\theta$ while the UAV altitude has great impact on the first moment. This is because the LoS links are much better than NLoS links and increasing the UAV altitude can increase the probabilities of establishing LoS links, and noise has greater impact at lower values of transmit power.

\begin{figure}[ht]
	\centering
	\subfigure[]{\includegraphics[width=0.8\columnwidth]{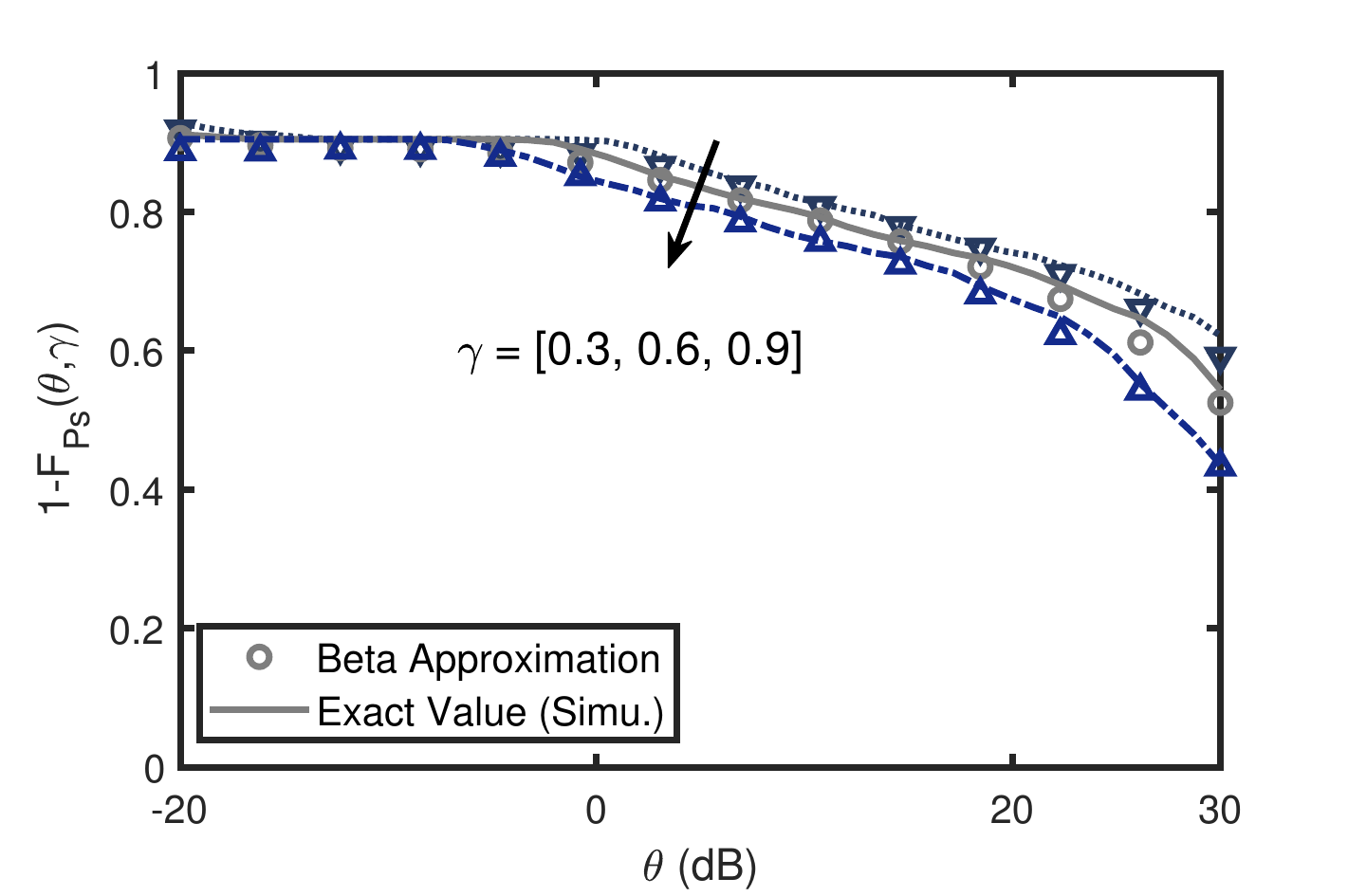}}
	\subfigure[]{\includegraphics[width=0.8\columnwidth]{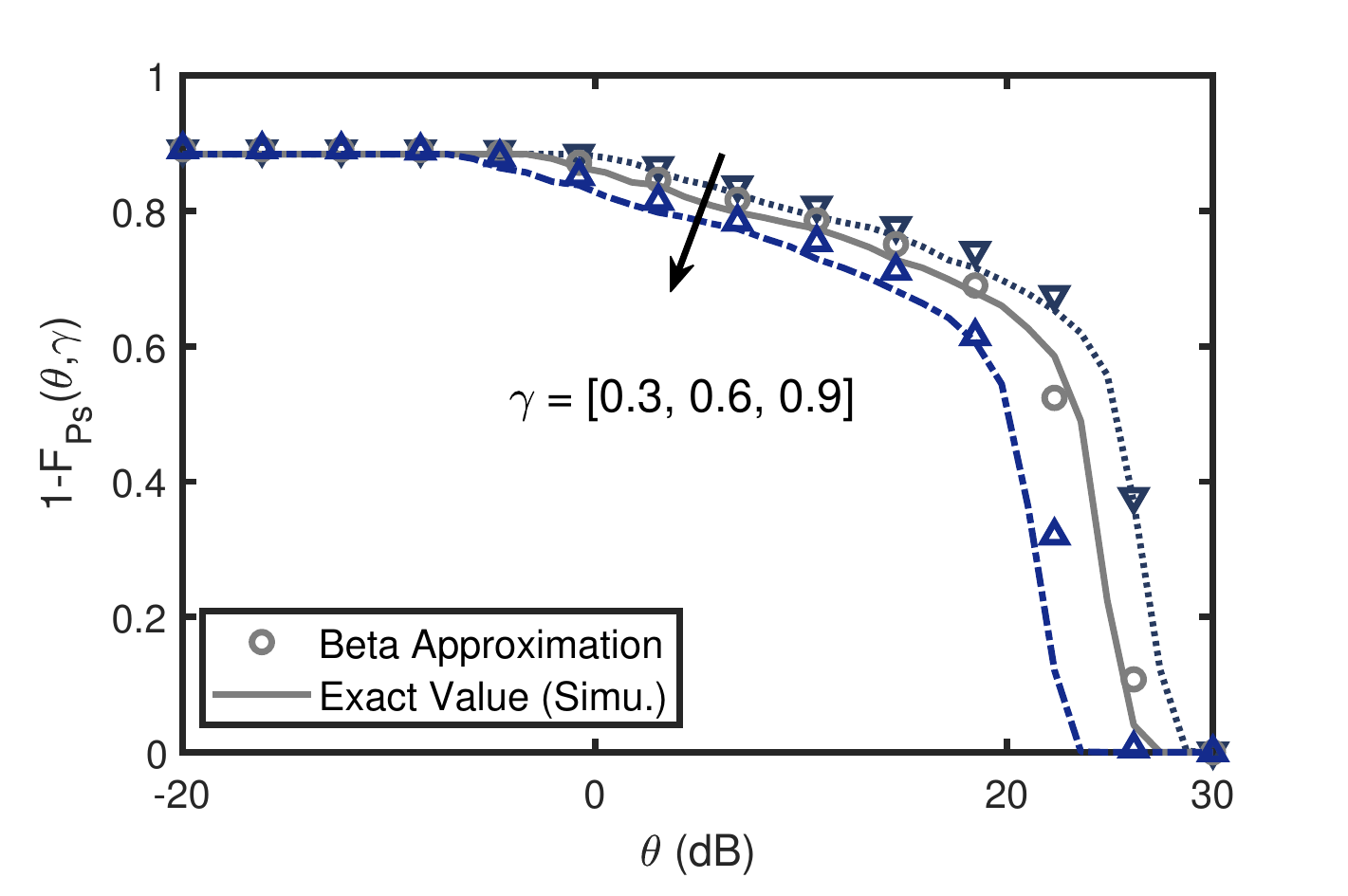}}
		\vspace{-2mm}
	\caption{The simulation and beta approximation results of the SINR meta distribution at \textbf{(a)} $\epsilon_{l,n} = \max$ and \textbf{(b)} $\epsilon_{l,n} = 0$, under different values of $\theta$ and $h = 100$ m.}
	\label{fig_Meta_DifEpsilon}
	\vspace{-2mm}
\end{figure}

\begin{figure}[ht]
	\centering
	\subfigure{\includegraphics[width=0.8\columnwidth]{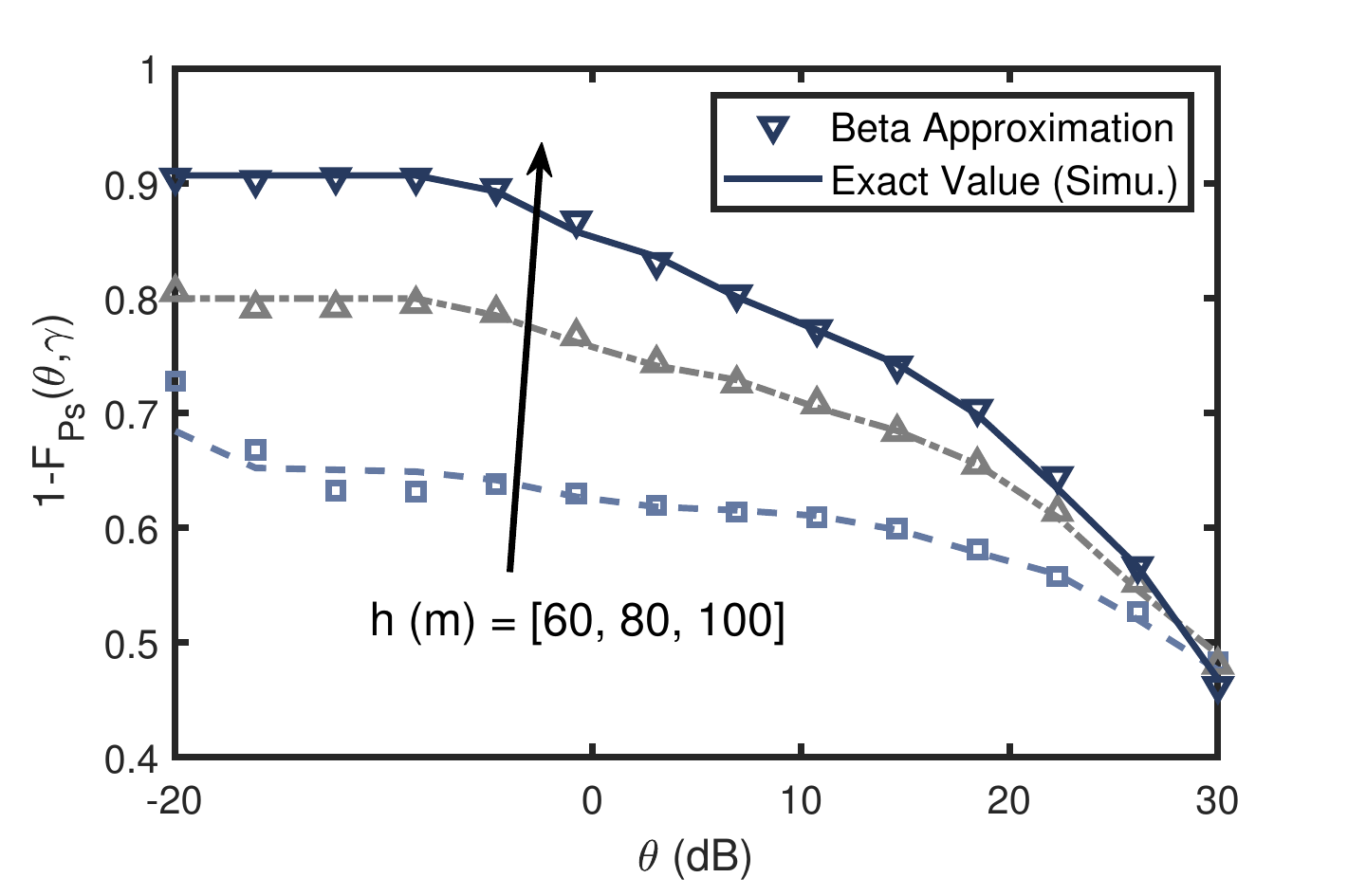}}
	\vspace{-2mm}
	\caption{The simulation and beta approximation results of the SINR meta distribution at UAV altitudes under different values of $\theta$, $\gamma = 0.9$ and $\epsilon_{\{l,n\}} = \max$.}
	\label{fig_Meta_DifH}
	\vspace{-2mm}
\end{figure} 
	\vspace{-1mm}

The differences between traditional coverage probability ($M_1(\theta)$ in Fig. \ref{fig_M1M2}) and SINR meta distribution can be observed by comparing Fig. \ref{fig_M1M2} and Fig. \ref{fig_Meta_DifEpsilon} and Fig. \ref{fig_Meta_DifH}. For instance, for a given $\theta$, while $M_1(\theta)$ is a constant, $\bar{F}_{P_s}(\theta)$ is a function of $\gamma$. That is, while coverage probability is the mean value of $P_s(\theta)$, SINR meta distribution is the CCDF of $P_s(\theta)$. In the case of two networks having the same coverage probability, the operators can distinguish them by comparing their CCDF.
	
 Fig. \ref{fig_Meta_DifEpsilon} and Fig. \ref{fig_Meta_DifH} provide more system insights about the effect of $\epsilon_{l,n}$ and UAV altitudes. In Fig. \ref{fig_Meta_DifEpsilon}, the SINR meta distribution does not decrease a lot with the increase of $\gamma$ at low values of $\theta$ ($\theta$ $<$ 20 dB). It implies that once a LoS link established between the user and the cluster UAV, this link can be of very high quality, e.g. high SINR. The sharp decrease of SINR meta distribution in Fig. \ref{fig_Meta_DifEpsilon}(b) is owing to the noise. If we ignore the difference in transmit power and let $G_{\{\cdot\}}$ denotes the channel fading and $x_{\{\cdot\}}$ denotes the locations of users, the SINR can be rewritten as $\frac{ G_{0}\left\|x_{0}\right\|^{-\alpha}}{\sum_{x_{i} \in \Phi_{u} \setminus x_{0}, i\in\mathbb{N}} G_{i} \|\left. x_{i}\right\|^{-\alpha}+\frac{\sigma^{2}}{p_t}}$ and the term $\frac{\sigma^{2}}{p_t}$ has higher impact on the conditional success probability at larger values of $\theta$ (e.g., success probability drops quickly at large values of $\theta$).

Fig. \ref{fig_Meta_DifH} shows the SINR meta distribution under different values of UAV altitudes. Compared to the impact of the compensation factor, system's reliability increases dramatically with the increase of UAV altitudes at low values of SINR threshold, say $\theta<20$ dB. This is because of the better performance of the LoS links. Users are more likely to establish LoS links with the cluster UAVs at higher altitudes. We also notice that UAV altitudes have higher impact on system's reliability at low values of $\theta$ ($\theta<20$ dB), and compensation factors have higher impact on large values of $\theta$ ($\theta>20$ dB).  Besides, we notice that the beta approximation does not perform well at low UAV altitudes.
	\vspace{-2mm}
\section{Conclusion}
	\vspace{-1mm}
This work studies the SINR meta distribution of UAV-enabled uplink wireless networks with inversion power control. Conditional success probability and moments are derived for computing the exact expression of SINR meta distribution and moment matching approximation of SINR meta distribution. Our numerical results show that the altitude of UAVs have higher impact on the system reliability than transmit power at low values of $\theta$ since users benefit more by establishing LoS links with UAVs.

\appendix

\subsection{Proof of Lemma \ref{lemma_ps}}\label{app_proof_ps}
Equations (\ref{eq_ps_l}) and (\ref{eq_ps_n}) are derived by
\begin{align}
	&\mathbb{P}(\frac{p_{t,l}\eta_{ l} G_l R_u^{-\alpha_l}}{I+\sigma^2}>\theta|{\rm  LoS}) =\mathbb{P}(G_l>\frac{\theta R_u^{(1-\epsilon_u)\alpha_l}}{\eta_{ l} \rho_l}(I+\sigma^2)|{\rm  LoS})\nonumber\\
	& \stackrel{(a)}{\approx}(1-\exp(-\beta_2(m_l)m_ls_l(R_u)))^{m_l}\nonumber\\
	&\stackrel{(b)}{=}\sum_{k = 1}^{m_l}\binom{m_l}{k}(-1)^{k+1}\exp(-\beta_2 (m_l)m_l ks_l(R_u)(I+\sigma^2)),\label{eq_proof_psl}
\end{align}
where $(a)$ follows from $s_l(r) = \theta r^{(1-\epsilon_{l})\alpha_{ l}}(\rho_l\eta_{ l})^{-1}$, the definition and the upper bound of Gamma distribution \cite{galkin2019stochastic,alzer1997some}: $\frac{\Gamma_{l}(m,mg)}{\Gamma(m)}<(1-\exp(-\beta_2(m)mg))^{m}$, where $\Gamma_{l}(m,mg)$ is the lower incomplete Gamma function \cite{bai2014coverage}, and $(b)$ results from the binomial theorem (assuming that $m_l$ and $m_n$ are integers). (\ref{eq_ps_n}) follows a similar steps, thus omitted here. The proof completes by substituting (\ref{eq_I}) into (\ref{eq_proof_psl}).

\subsection{The Laplace Transform in Lemma \ref{lemm_Mb}}\label{app_laplace}
The expression of the Laplace transform in Lemma \ref{lemm_Mb} is given by
	\begin{align}
		&\mathcal{L}(g_{1}(r),\cdots,g_{b}(r)) =  \exp(-(g_{1}(R_u)+\cdots+g_{b}(R_u))\sigma^2)\nonumber\\
		& \exp\bigg(-2\pi\lambda_u\int_{h}^{\infty}\int_{h}^{\sqrt{h^2+r_c^2}}\bigg[1-f(e,j,g_{1}(R_u),R_{u_i},D_{u_i}) \nonumber\\
		&\cdots\times f(e,j,g_{b}(R_u),R_{u_i},D_{u_i})\bigg] z P_e(\sqrt{r^2-h^2}) \nonumber\\
		&\times P_j(\sqrt{z^2-h^2})f_{\rm R_u}(r){\rm d}r{\rm d}z\bigg),\quad \{e,j\}\in\{l,n\}.
	\end{align}

\bibliographystyle{IEEEtran}
\bibliography{Ref9}
\end{document}